# RecA-mediated homology search as a nearly optimal signal detection system

Yonatan Savir[1] and Tsvi Tlusty[1,*]

[1] *Department of Physics of Complex Systems, Weizmann Institute of Science, Rehovot, 76100, Israel.*
*Correspondence: tsvi.tlusty@weizmann.ac.il

**SUMMARY**

**Homologous recombination facilitates the exchange of genetic material between homologous DNA molecules. This crucial process requires detecting a specific homologous DNA sequence within a huge variety of heterologous sequences. The detection is mediated by RecA in E. coli, or members of its superfamily in other organisms. Here we examine how well is the RecA-DNA interaction adjusted to its task. By formulating the DNA recognition process as a signal detection problem, we find the optimal value of binding energy that maximizes the ability to detect homologous sequences. We show that the experimentally observed binding energy is nearly optimal. This implies that the RecA-induced deformation and the binding energetics are fine-tuned to ensure optimal sequence detection. Our analysis suggests a possible role for DNA extension by RecA, in which deformation enhances detection. The present signal detection approach provides a general recipe for testing the optimality of other molecular recognition systems.**

**INTRODUCTION**

Homologous recombination (HR) enables the organism to exchange genetic information between two almost identical DNA strands, a crucial process for maintaining the integrity of the genome as well as for generating genetic diversity. Successful recombination requires the efficient detection of a target DNA within a vast pool of heterologous competitors. The proteins of the RecA superfamily, which includes RecA, RadA, Rad51 and DMC1, have been identified as the recombinases that can activate HR. The conservation of structure and function throughout this family of recombinases allows the use of bacterial RecA as a model for protein-mediated HR (Bianco et al., 1998; Roca and Cox, 1997).

The multistage process of HR (Kowalczykowski, 1991b; Kuzminov, 1999; Lusetti and Cox, 2002; Radding et al., 1983) includes the formation of a RecA-DNA filament, the search for homologous DNA sequences, the formation of a synapse, and the exchange of strands (Fig. 1). In the first stage, RecA monomers polymerize along a single-stranded DNA (ssDNA) and form a nucleoprotein filament (NPF) (Fig. 1AB). Inside the NPF, the ssDNA is in a highly deformed conformation. The DNA is stretched and underwound, with an average rise of about 5.1Å per nucleotide and 18.5 nucleotides per turn (Chen et al., 2008; Flory et al., 1984; Stasiak and Di Capua, 1982). The deformation of the ssDNA inside the NPF is non-uniform and its structure exhibits clear 3-base periodicity (Chen et al., 2008): Each nucleotide triplet itself is only slightly deformed and maintains a B-DNA-like conformation with an average rise of 4.2 Å. Most of the deformation is localized between the triplets such that the axial rise between the bases at the edges of each triplet is increased from 3.4 Å to about 7.8 Å.

The next stage, which is the subject of the present study, includes searching for homology and strand exchange between the NPF and double stranded DNA (dsDNA) molecules in its vicinity (Fig. 1CD). This process does not simultaneously engage the entire molecules but instead occurs in a *stepwise* fashion (Cox, 2007). Homology check involves the initial alignment between a short region of the NPF and a corresponding short segment along the dsDNA, which binds to the secondary binding site of the NPF (Mazin and Kowalczykowski, 1998) to form a three-stranded synaptic intermediate (later, the same binding site binds the displaced ssDNA). The dsDNA segment must partly melt and deform non-uniformly in order to align its base pairs with the bases of the RecA-polymerized ssDNA and to allow recognition via Watson-Crick pairing (Chen et al., 2008; Folta-Stogniew et al., 2004; Lee et al., 2006; Voloshin and Camerini-Otero, 2004; Zhou and Adzuma, 1997). Since each triplet of bases in the RecA-polymerized ssDNA is similar to the B-DNA structure, the triplets of the dsDNA are not significantly deformed, but for the next triplet to align with the ssDNA bases, the dsDNA has to locally extend by a factor of almost 2.5 (from 3.4Å to about 8Å) (Chen et al., 2008). After the bases are aligned, a 'decision' is made. If the segments are heterologous, the synapse is destabilized and may disintegrate. If they are homologous, the synapse is stable and the process may proceed by checking a subsequent segment for homology. Strand invasion occurs at one end of the synapse, which is about 80 bp long (van der Heijden et al., 2008), whereas at the other end, RecA unbinds (Fig. 1E).

Accumulating evidence suggests that each HR step involves only a small number of base pairs. It has been shown that the decision as to whether the sequences are homologous or not is made before the synapse reaches a length of about 20-30 bp (Sagi et al., 2006; Shen and Huang, 1986). This minimum efficient processing segment (MEPS) sets an upper limit to the size of each step. Examination of the effect of mismatches on HR mediated by bacterial RecA and yeast Rad51 (Bucka and Stasiak, 2001; Holmes et al., 2001) suggests a homology check step of a few bp (about 3-10). This result is in accordance with studies of base pair flipping during HR (Folta-Stogniew et al., 2004; Voloshin and Camerini-Otero, 2004), which imply that the homology check is mediated via partial melting of a few base pairs, and with theoretical models of HR (Dorfman et al., 2004). The recently discovered 3-base periodicity (Chen et al., 2008) gives further support to the suggestion that the homology check process involves short steps with a natural length scale of a base triplet (Kowalczykowski, 2008).

Although ATP hydrolysis is required for RecA depolymerization during HR and affects its efficiency, homologous search is known to occur even in the absence of ATP hydrolysis (Menetski et al., 1990; Rice et al., 2001; Rosselli and Stasiak, 1990, 1991; Xiao et al., 2006). Similarly, in the eukaryotic Rad51 system, NPF formation, homology search and even DNA strand exchange do not require ATP hydrolysis (Rice et al., 2001; Sung and Stratton, 1996). This implies that the process of homologous search can be carried out solely by diffusion and equilibrium thermal fluctuations.

The RecA-polymerized ssDNA has to cope with a non-trivial task: it needs to detect its complementary sequence along a dsDNA that contains many heterologous sequences. The task of homologous recognition is therefore a specific case of a general *signal detection problem* in which a system has to detect a correct signal among various competing false signals. To test whether RecA has evolved to optimally solve this problem, we used signal detection theory to construct a measure for the quality of sequence detection as a function of the binding energetics. We show that the measured interaction energies nearly maximize the quality of the detection process. That is, there is a nearly optimal tuning of the binding and deformation energies. We also sketch possible experimental directions that can test this suggestion. Our analysis provides alternative explanation for the extension of DNA during HR in which deformation is necessary in order to enhance detection (Savir and Tlusty, 2007, 2008).

## RESULTS

### Homologous recognition as a signal detection problem

Signal detection theory deals with decision making under uncertainty and provides quantitative measures to estimate the quality of the decision. We use this framework to derive a measure for the discrimination ability of molecular recognizers. Though we focus on HR, other recognition process can also be analyzed using this framework (Savir and Tlusty, 2008).

In general, a decision unit receives at its input a signal and, in response, outputs a decision. In molecular recognition systems, the input signals are often noisy and the decision process is prone to errors. In the case of the HR system, each step of the homologous search is a decision-making process whose possible input-output combinations are summarized in Fig. 2: the possible inputs are the encounters of the NPF with a complementary dsDNA segment (denoted by a '+') or a non-complementary one ('−'). The possible outcomes are whether to proceed with the homologous recombination process to the next step ("go" denoted by a '+') or to abort it ("no go", '−'). Each of the four input-output combinations occurs with a certain probability and has a different impact on the system.

A natural measure for the quality of detection is the average impact on the system (the average Bayesian cost), $\sum_{i,j} C_{ij} p_{ij}$, where $p_{ij}$ is the probability of receiving an input $j$ and making a decision $i$, and the weight $C_{ij}$ measures the consequences of this decision (Helstrom, 1995). For example, in some systems interpreting a '−' signal as a '+' ("false alarm") may be lethal and the corresponding weight $C_{+,-}$ will be very large, whereas in other systems it may have only a minute effect and the weight will be small. To optimize the quality of detection, one must maximize the benefit owing to correct decisions while minimizing the penalty for making incorrect ones. These benefits and penalties are reflected in the opposite signs of the weights: negative weights for correct decisions and positive weights for incorrect ones. Minimizing the cost with respect to the various system parameters yields the optimal Bayesian decision rule which resolves the tradeoff between maximizing the benefits and minimizing the penalties.

At the molecular level, the NPF encounters as an input a complementary or a non-complementary dsDNA segment with probabilities, $p_h(+)$ and $p_h(-)$, respectively. The decision step involves the binding of an NPF segment to the corresponding dsDNA segment at a probability $p_b$. The probability that the formed complex is functional and that the HR process follows is $p_f$ ($p_f(+)$ for a functional complementary complex and $p_f(-)$ for a functional non-complementary complex). The average Bayesian cost function (SI) takes the form

$$C_b = -c_+ \cdot p_h(+) \cdot p_b(+) \cdot p_f(+) + c_- \cdot p_h(-) \cdot p_b(-) \cdot p_f(-), \quad (1)$$

where $c_+ = C_{-,+} - C_{+,+}$ and $c_- = C_{+,-} - C_{-,-}$ are both positive. The first term of (1) is proportional to the rate of correct HR, whereas the second term is proportional to the rate of incorrect HR.

### Detection quality depends on RecA-DNA energetics

Next, we examine the effect of binding energetics on the detection quality. Motivated by the experimental evidence reviewed above, we modeled the initial recognition event as follows (Fig. 1C). Each *step* of the homology check involves a few base pairs, $N$. In each step, a dsDNA segment of $N$ bp has to stretch (Fig. 1C), non-uniformly, in order to align with the polymerized ssDNA. This stretching requires extension free energy, $\Delta G_{ext}$ per triplet (within the triplet there is a slight deformation). The average free energy gained from pairing one of the strands of the dsDNA and the polymerized ssDNA, which includes the effect of the secondary binding site of the RecA, is $\Delta G_b$ per bp. The overall average free energy gain per bp is $\Delta G_t$ (which depends on $\Delta G_b$ and $\Delta G_{ext}$, see next section).

We denote by $m$ the number of base pair mismatches between the two segments (homologous, complementary segments have $m = 0$). The probability that the interacting segments have $N - m$ complementary bp is $p_h(N,m)$. If the two segments are complementary, the initial complex is likely to be stabilized and the process of HR may continue. Otherwise, this recognition nucleus may dissociate. This is reflected in the binding probability between the segments, $p_b$, and in the probability that the formed complex is functional, $p_f$. Clearly, both probabilities, $p_b$ and $p_f$, depend on $m$ and are maximal for complementary strands ($m = 0$).

Since homologous search takes place even in the absence of ATP hydrolysis (see Introduction, (Menetski et al., 1990; Rosselli and Stasiak, 1990, 1991; Xiao et al., 2006)), the binding between the corresponding complementary segments of the NPF and the dsDNA segment is estimated by the equilibrium binding probability, $p_b(+) = 1/[1 + \exp(-N \cdot \Delta G_t)]$, where $\Delta G_t$ is the overall average free energy gain of the step per bp, measured in units of $k_B T$. If the interacting segments are non-complementary, i.e., there are $m > 0$ base mismatches, the binding free energy is reduced to $p_b(-) = 1/[1+\exp(-N \cdot \Delta G_t + m \cdot \Delta\Delta G_{mis})]$, where $\Delta\Delta G_{mis}$ is the average free energy loss due to one mismatch (Malkov and Camerini-Otero, 1998; Malkov et al., 1997). It is assumed that dsDNA has to stretch (non-uniformly) in order to interact properly with the RecA-polymerized ssDNA. No other assumptions need to be made regarding the exact nature of the specific interaction or the nature of the stable complex. As we will discuss next, the free energies, $\Delta G_t$ and $\Delta\Delta G_{mis}$, can be extracted from the available experimental data, which allows us to estimate the actual cost function.

We now consider the case in which the polymerized ssDNA has to discern between a complementary dsDNA segment with no mismatches, $m = 0$, and a dsDNA with $m \geq 1$ mismatches. By introducing the binding probabilities into $C_b$ (1), we derive a *normalized* measure $C$ (SI text) for the quality of the homologous recognition step as a function of the interaction energies, the number of bp involved in the process, $N$, and the number of mismatches, $m$,

$$C = \frac{-1}{1+\exp(-N \cdot \Delta G_t)} + \frac{(1/t)}{1+\exp(-N \cdot \Delta G_t + m \cdot \Delta\Delta G_{mis})}. \quad (2)$$

The first term represents the benefit of detecting a complementary sequence, whereas the second term represents the penalty for detecting a non-complementary sequence. The optimal molecular detector must balance the two terms of (2) in order to minimize the overall detection cost. The parameter $t = [c_+ \cdot p_h(+) \cdot p_f(+)] / [c_- \cdot p_h(-) \cdot p_f(-)]$ represents the *tolerance* of the system. As $t$ decreases, the penalty for detecting a non-complementary sequence increases and the system is less tolerant of errors. For example, this may occur when the penalty for an incorrect decision, $c_-$, is much higher than the benefit of a correct one, $c_+$, or when non-complementary sequences are much more abundant than homologous ones. Similarly, an error-tolerant system is characterized by a higher value of $t$. In the case of HR, a rough estimate of $t$ is $t \approx 1$ (see SI).

**Optimal homology detection**

Our aim is to estimate the total binding free energy, $\Delta G_t$, which is optimal for homology recognition, (i.e. minimizes the detection cost (2)). Evidently, if $\Delta G_t$ increases, then the probability of correct binding in (2) increases as well. However, at the same time, the probability of incorrect binding also increases (Fig. 3A). At the other extreme, very small $\Delta G_t$ values may reduce the incorrect binding, but they also reduce the correct ones. The optimal value of $\Delta G_t$ decreases the binding probability of both correct and incorrect sequences, but the reduction of the incorrect sequences is larger and thus the overall detection improves. For example, in the symmetric case where the impact of correct and incorrect binding is similar, $t = 1$, we find that the optimal free energy per bp, $\overline{\Delta G_t}$, is (Fig. 3A)

$$\overline{\Delta G_t} = \frac{m}{2N} \cdot \Delta\Delta G_{mis} . \qquad (3)$$

The solution for general values of $t$ is straightforward (see Methods, Fig 3).

To test whether DNA-RecA energetics are optimal, we estimate the experimental values of $\Delta G_t$ and $\Delta\Delta G_{mis}$. Measurements of HR kinetics (Lee et al., 2006; Malkov and Camerini-Otero, 1998; Malkov et al., 1997; Xiao et al., 2006; Yancey-Wrona and Camerini-Otero, 1995) imply that $\Delta G_t$ is of the order of 1 $k_BT$ (see Methods, SI). This value was previously used in theoretical models of HR (Dorfman et al., 2004; Fulconis et al., 2005). The destabilization energy, $\Delta\Delta G_{mis}$, was measured (Malkov and Camerini-Otero, 1998; Malkov et al., 1997) to be in the range of 1.5 − 3.5 $k_BT$, which is consistent with a recent study (Volodin et al., 2009) (see Methods). Following the experimental evidence, especially the recently discovered 3-base periodicity (Chen et al., 2008), we assume a step size of length 3, the size of one RecA monomer, so each step involves one inter-triplet stretch.

The resulting cost curves, $C(\Delta G_t)$, are graphed in Figure 3B. The possible competitors in this case are segments with $m = 1, 2,$ or 3 mismatches. There is a well-defined 'valley' where the optimal values of $\Delta G_t$ lie. The resulting "optimum band" lies within the range of observed $\Delta G_t$ values. Competitors with more mismatches correspond to higher optimal $\Delta G_t$ values, since the potential competitor is less attractive and increasing the binding energy does not compromise detection.

Since an exact value of the tolerance $t$ is unavailable, and to account for the possibility of different $t$ values of different competitors, we examine the sensitivity of the results to variations in $t$ (Fig. 3C). The priority of a system with low error-tolerance, $t < 1$, is to minimize the incorrect binding at the expense of correct binding. In such a system, the penalty for binding a non-complementary sequence is higher. High $\Delta G_t$ values are therefore disadvantageous, since they allow more incorrect binding. As a result, the optimal binding energy is slightly shifted to lower $\Delta G_t$ values. Likewise, at high error-tolerance, $t > 1$, the optimal free energy $\Delta G_t$ is shifted to slightly higher values.

In the limiting case of a detection system having very low error-tolerance (Fig. 3C), $t < t_c = \exp(-m \cdot \Delta\Delta G_{mis})$, the cost $C$ is governed by the need to avoid errors (the second term of (2)). In this case, the cost function has a sharp sigmoidal shape, a kin to a step function, with a threshold at $\Delta G_t \approx (m/N) \cdot \Delta\Delta G_{mis}$. In the lower part of the sigmoid, where the cost is practically constant, larger binding energies allow the same sequence discrimination at a higher rate. The largest binding energy that still resides in the lower part of the sigmoid is near the transition point. Therefore, even at the extreme of very low tolerance, free energy values of about 1 $k_BT$ are functionally advantageous.

So far, we did not discuss the actual base pair content of the segment but only the number of mismatches. However, the AT interaction is weaker than the GC interaction and indeed HR is known to take place first in AT-rich regions (Gupta et al., 1999). The differences in $\Delta\Delta G_{mis}$ between AT and GC pairs is about ~ 1 $k_BT$ (Malkov and Camerini-Otero, 1998). Taking this into account, we find that a GC-rich region corresponds to an optimal $\Delta G_t$ larger by ~ 0.2 $k_BT$ than that of an AT-rich region. We also examined the sensitivity of our results to variations in $N$. For the estimated range of recognition step size, $N = 3 −10$, taking into account the various competitors, $1 \leq m \leq N$, we found (SI) that the optimal free energy still falls within the range of the measured values.

**Deformation and binding are optimally tuned**

The role of the RecA-induced ssDNA extension, which leads to a geometrical mismatch between the polymerized ssDNA and the dsDNA, is not fully understood. It was suggested that this deformation enables homology check by destabilizing the dsDNA (Chen et al., 2008) or may prevent topological traps (Klapstein et al., 2004). We derive the conditions in which the optimal deformation is non-zero. We thus suggest an alternative explanation, in which deformation enhances the detection capacity of the HR process.

The total free energy gain $\Delta G_t$ depends on the extension free energy, $\Delta G_{ext}$, and on the binding free energy, $\Delta G_b$. For a step size 3 bp long, $N = 3$, each step involves one extension and $\Delta G_t = \Delta G_b - \Delta G_{ext}/3$. The factor of 1/3 arises because $\Delta G_{ext}$ is given per triplet (see SI for general $N$). Since the measured value of $\Delta G_t$ is about 1 $k_B T$, the extension and binding free energies are limited to certain values such that $\Delta G_{ext} \approx 3 \cdot (\Delta G_b - 1)$ (Fig. 4). A direct result of the optimality of $\Delta G_t$ (3) is a relation between the optimal values $\overline{\Delta G_{ext}}$ and $\overline{\Delta G_b}$,

$$\overline{\Delta G_{ext}} = 3\left(\overline{\Delta G_b} - \frac{m}{6} \cdot \Delta\Delta G_{mis}\right). \tag{4}$$

Figure 4 (inset) graphs the experimental and the theoretical limitations on $\Delta G_{ext}$ and $\Delta G_b$. The resulting bands are parallel and overlapping. This, of course, follows directly from the (near-) optimality of the total free energy.

Figure 4 demonstrates the impact of the extension free energy $\Delta G_{ext}$ on the binding probabilities and on the detection cost (Fig. 4A). An increase in deformation may reduce the incorrect binding but also reduce the correct ones. The optimal deformation is such that a complementary sequence tends to stretch, so it may gain binding energy by interacting with the deformed ssDNA inside the NPF, whereas the non-complementary sequence, owing to its lower affinity, is less likely to stretch since it gains less binding energy.

Relation (4) implies that binding and extension co-evolved such that their values optimally match. We cannot determine whether the extension was evolved to match the binding energies or vice-versa. However, as long as $\Delta G_b > (m/6) \cdot \Delta\Delta G_{mis} \approx 1$ $k_B T$, the optimal extension is non-zero. If, due to other constraints of the system, for example optimizing kinetic rates, $\Delta G_b$ is required to be larger than 1 $k_B T$, then RecA-mediated extension of the DNA enhances the detection efficiency of HR.

**DISCUSSION**

Homologous recombination requires the detection of a specific sequence within a variety of structurally similar competitors. A reasonable hypothesis that we examined in the present work is that the energetic and structural features of RecA have evolved to efficiently solve this detection problem. The analogy to signal detection allowed us to use a natural cost measure to examine the full functional tradeoff of the system: maximizing the correct decisions, 'positive design', while minimizing the incorrect ones, 'negative design'. Indeed, according to our analysis, the observed overall free energy, $\Delta G_t$, optimizes this tradeoff. This suggests that the observed deformation and binding energies are optimally tuned.

In principle, our quantitative predictions can be experimentally tested using existing techniques. Below we sketch a few examples for such possible tests. For example, in a single-molecule experiment, a dsDNA fixed to a surface is placed in a solution containing RecA and homologous or heterologous ssDNA sequences. The ssDNA and RecA will form NPFs, which will recombine with the fixed dsDNA. One then can externally control the extension and helicity of the dsDNA, for example, using magnetic tweezers (Fulconis et al., 2006), and measure their effect on the resulting fraction of exchange (FOE). During the homologous search, the dsDNA has to deform non-uniformly in order to be in registry with the bases of the polymerized ssDNA (Chen et al., 2008) (Fig. 1). Thus, stretching the DNA, say by a factor of 5-10%, reduces the overall required deformation and, as a result, the overall $\Delta G_t$ will increase. Owing to the increase in the free energy gain, our model predicts that both homologous and heterologous FOEs will increase with the applied external stretch, but the detection quality, the difference between the FOEs, is predicted to *reduce*. Another direction utilizes RecA mutants with either reduced or enhanced strand exchange activity (Kowalczykowski, 1991a; McGrew and Knight, 2003). In both cases, our model predicts that the detection cost will be reduced since deformation and binding are not tuned anymore.

By and large, molecular recognition systems involve intricate structural and energetic details that are often lacking or are measured only at low resolution. In such cases – when several models are suggested or when a suggested model depends on many unknown parameters – our signal detection framework may be useful in selecting those models that have functional advantage. The selected models should obey the predicted optimality relations, such as (3-4), between the structural parameters (e.g., recognition step size, extension energy, the gap between DNA triplets) and biochemical parameters (e.g., affinity).

One can apply this approach to other recombination systems. For example, high resolution structure of human Rad51 bound to DNA is still lacking. A recently proposed model (Reymer et al., 2009) relies on combining known domains of hRad51, low resolution structure of yeast Rad51 reconstructed from electron micrographs and DNA conformation in RecA-DNA complexes (Chen et al., 2008). This model predicts, as was observed in RecA, that the DNA is extended in a non-uniform manner with gaps between nucleotide triplets. Alternatively, instead of relying on RecA-DNA structure, our model can estimate the optimal distance between consecutive triplets based on Rad51 binding energies. Such estimates can be used to refine low-resolution structural models (Shin et al., 2003) or to assess high-resolution models using solely Rad51 data.

In our model, introducing a dsDNA extension may improve the overall performance of the detection (Fig. 4A). This suggests a general design principle, termed *conformational proofreading*, in which introducing a structural mismatch between the recognizer and its target enhances the quality of detection (Savir and Tlusty, 2007, 2008). It is instructive to compare the suggested design principle to kinetic proofreading (Hopfield, 1974; Ninio, 1975) in which introducing a time delay (or an additional irreversible stage) in the formation of correct and incorrect complexes reduces the production rates but enhances the fidelity beyond the equilibrium limit. The time delay in kinetic proofreading is therefore analogous to the spatial deformation in conformational proofreading. However, the conformational proofreading is an *equilibrium* scheme that does not consume energy. We suggest that the HR process, which is carried out while ATP is being hydrolyzed, consists of both mechanisms. The initial recognition includes a step of conformational proofreading that involves deformation upon binding, followed by a possible kinetic proofreading step that utilizes ATP hydrolysis (Bar-Ziv et al., 2002; Sagi et al., 2006; Tlusty et al., 2004).

The fact that a similar helical structure of nucleo-protein filaments was conserved among various proteins in the RecA superfamily across phylogenetically distant organisms (Egelman, 2001) suggests that these structures have an underlying common functional advantage. The present work reveals such an advantage: the energetics of binding and stretching the dsDNA during HR is fine-tuned together with the RecA-induced extension of ssDNA to optimize the quality of detection. Our signal detection approach suggests a general recipe for testing the optimality (Savir et al., 2010) of other molecular recognition systems.

**METHODS**
**Optimal free energy.**
We obtain the optimal binding free energy, $\Delta G_t$, by minimizing the detection cost function $C$. A straightforward derivation yields,

$$\overline{\Delta G_t} = N^{-1} \ln\left[ (1-t \cdot t_c)^{-1} \cdot \left( (t-1) + \sqrt{(t/t_c) \cdot (t_c-1)^2} \right) \right] \quad (5)$$

where $t_c = \exp(-m \cdot \Delta\Delta G_{mis})$. It follows that for $t = 1$, $\Delta G_t = (m/2N) \cdot \Delta\Delta G_{mis}$ (3). As the tolerance to errors $t$ is reduced, the optimal free energy increases. If the system has a very low tolerance, $t \leq t_c$, the detection cost does not exhibit a minimum but instead displays a sharp sigmoid. In this case, the inflection point of this sigmoid is $\Delta G_t \approx (m/N)\Delta\Delta G_{mis}$.

**Experimental values of the binding free energies**
The destabilization energy resulting from one mismatch, $\Delta\Delta G_{mis}$, was measured by introducing mismatches into RecA-mediated HR in the absence of ATP hydrolysis (Malkov and Camerini-Otero, 1998; Malkov et al., 1997). These experiments yielded an estimate for $\Delta\Delta G_{mis}$ of about $1.5 - 3.5$ $k_B T$, which was found to be consistent with a recent study in which HR of short oligonucleotides was measured (Volodin et al., 2009).

In a recent series of studies (Lee et al., 2006; Xiao et al., 2006), the kinetics of strand exchange in the presence of ATPγS was measured at different temperatures using fluorescence techniques. In a typical experiment, a 30-nt ssDNA marked with a fluorophore is pre-incubated with RecA to form a nucleo-protein filament (NPF). This NPF is then mixed with a homologous dsDNA. By fitting a kinetic model to the temporal dependence of the fluorescence intensity, the kinetic parameters of the process were extracted, from which the energetics of the process were evaluated. During the reaction, the dsDNA locally melts, stretches, and one strand is paired with its complementary RecA-polymerized ssDNA. The thermodynamic measurement of the progression of this reaction is graphed in figure S1. The formation of a synapse is manifested by two intermediates (denoted by the authors as $N_2$ and $N_3$) in which all three strands are in an extended, stretched conformation where the incoming ssDNA is paired with its complementary counterpart. By a straightforward analysis of their results (see SI), one finds that the free energy gain involved in forming the synapse is 30 $k_B T$ per segment in the $N_2$ intermediate, i.e., 1 $k_B T$ per bp, and 45 $k_B T$ per segment in the $N_3$ intermediate, i.e., 1.5 $k_B T$ per bp. An estimate for the overall binding free energy is therefore -1 1.5 $k_B T$. The final product of this reaction involves strand displacement that is not part of the synapse formation (the strand is not released due to the absence of ATP). The free energy of this final product is 60 $k_B T$ per segment, or 2 $k_B T$ per bp, which therefore sets an upper bound on $\Delta G_t$.

The dissociation constant for pairing between a RecA-polymerized ssDNA of 24-27 bp and dsDNA was found to be around 80-130nM (Yancey-Wrona and Camerini-Otero, 1995), which corresponds to $\Delta G_t$ of around 0.7 $k_B T$ per bp. For comparison, the dissociation constant for pairing in the presence of ATP hydrolysis was found to be around 200 nM (Bazemore et al., 1997). In another study (Malkov et al., 1997), the free energy that accompanies the formation of a complex between an NPF and a supercoiled dsDNA bearing a partly homologous region of 27 bp was evaluated. This study estimated that the destabilization free energy, $\Delta\Delta G_{mis}$ and the destabilization free energy was in the range of, is around 3 $k_B T$. The evaluated ratio between the overall free energy of homologous complexes and the destabilization free energy was in the range of 1.5 − 3 (i.e., $N \cdot \Delta G_t \approx 2.2 \cdot \Delta\Delta G_{mis}$), which yields a rather low estimate for $\Delta G_t$, of less than 0.5 $k_B T$.

## Acknowledgments

We thank Elisha Mosses, Adam Many, Asaf Tal and Rinat Goren for valuable discussions, and the support of the Israel Science Foundation grant 1329/08.

**Figure Legends**

Fig. 1: **Homologous recombination mediated by bacterial RecA**.
**(A)** RecA monomers polymerize along an ssDNA to form a nucleoprotein filament (NPF).
**(B)** The ssDNA inside the NPF is deformed non-uniformly (PDB 3CMU (Chen et al., 2008), RecA protein is not shown). Each nucleotide triplet has a B-DNA like structure. Most of the deformation is localized between the triplets such that the axial rise between neighboring triplets is increased from 3.4 Å to about 7.8 Å.
**(C)** The NPF searches for homology along a dsDNA. The homology check involves short segments in a stepwise fashion. The dsDNA deforms non-uniformly in order to align its base pairs with those of the RecA-polymerized ssDNA and thus allows recognition via Watson-Crick pairing. Therefore, the triplets of the dsDNA are slightly deformed, but for the next triplet to align with the ssDNA bases, the dsDNA has to locally extend by a factor of almost 2.5 (from 3.4 Å to about 8 Å). The aligned base pairs can be complementary (green rectangles) or non-complementary (green and red rectangles). Once the bases are aligned, a 'decision' is made. If the segments are heterologous, the synapse is destabilized and may disintegrate. If they are homologous, the synapse is stable and the process may proceed by checking a subsequent segment for homology.
**(D)** The paired DNA within the RecA filament exhibits 3-base periodicity. Most of the deformation is localized between the slightly deformed triplets (PDB 3CMT (Chen et al., 2008), RecA protein is not shown).
**(E)** Homology check and strand exchange occur at one end of the synapse. At the other end, RecA depolymerizes and allows the displacement of the exchanged ssDNA.

Fig. 2: **Decision table and tree in homologous recombination.** The inputs are whether the RecA-polymerized ssDNA (NPF) encounters a complementary dsDNA segment (+) or a non-complementary one (−). The outcomes are whether to proceed with HR (+) or to abort it (−). There are four input-output combinations: the correct True Positive (+, +) and True Negative (−, −), and the incorrect False Positive (+, −), i.e., 'false alarm', and False Negative (−, +), i.e., 'miss'. $p_{ij}$ is the probability of receiving an input, $j$, and making a decision, $i$. The weight, $C_{ij}$, measures the impact of such a decision on the system. Other recognition processes can be described by a similar Decision table.

Fig. 3: **RecA-DNA interaction is nearly optimal for sequence detection.**
**(A)** *Top*: As the total binding free energy per bp $\Delta G_t$ increases, the binding probability to a complementary sequence, $p_b(+)$ (blue), increases faster than the binding probability to a non-complementary one, $p_b(−)$ (red). *Bottom*: The detection cost $C$ (2) is the weighted difference between these two binding curves (green). Minimizing the detection cost amounts to maximizing the binding probability to a complementary segment while minimizing the binding probability to a non-complementary one. The optimal free energy is $\overline{\Delta G_t} = (m/2N) \cdot \Delta\Delta G_{mis}$ (plotted for tolerance $t$ =1).
**(B)** The detection cost $C$ (2) for a recognition step 3 bp long, $N = 3$, as a function of $\Delta G_t$ for $t = 1$. $C$ was calculated for all possible competitors, of $m = 1$ (green), 2 (red), and 3 (blue) mismatches. The cost function $C$ for each competitor appears as a band whose width is determined by the experimental range of the destabilization free energy $\Delta\Delta G_{mis} = 1.5 − 3.5\ k_B T$. The central value for each of the bands is denoted by a solid line. The shading denotes the experimental range of $\Delta G_t = 0.5 − 1.5\ k_B T$. For all competitors, the measured free energy is near the minimum of the cost curves and therefore enables nearly optimal detection.
**(C)** The detection cost for various tolerance values, $t$: the symmetric case, $t = 1$ (blue), low error-tolerance, $t < 1$ (red), high error-tolerance, $t > 1$ (green) and extremely low error-tolerance, $t < t_c = \exp(−m \cdot \Delta\Delta G_{mis})$ (cyan, normalized by $t$). In this figure, $m = 2$, $\Delta\Delta G_{mis} = 2.5\ k_B T$ and $N = 3$. In the case of HR, a rough estimate of $t$ is $t \approx 1$ (SI). At low error-tolerance, $t < 1$, avoiding incorrect binding is of higher priority than binding correctly and therefore low $\Delta G_t$ values are more beneficial relative to the symmetric case. The opposite phenomenon occurs at high error-tolerance (see text). For systems of extremely low error-tolerance, the cost has a sharp sigmoidal shape with a threshold at $\Delta G_t = (m/N) \cdot \Delta\Delta G_{mis}$. The largest binding energy that still resides in the lower part of the sigmoid is near the transition point. Therefore, even in at the extreme where the tolerance is very low, free energy values of about 1 $k_B T$ are advantageous (see text).

Fig. 4: **Binding and extension are tuned.** The binding probabilities, $p_b(+)$ (blue) and $p_b(-)$ (red), and the detection cost $C$ (green) as a function of the extension free energy per triplet, $\Delta G_{ext}$. The recognition step is 3 bp long. As the extension increases, the binding probabilities decrease such that $p_b(+)$ decreases slower than $p_b(-)$. Thus, the value of extension free energy that balances the opposite trends and minimizes the detection cost is $\overline{\Delta G_{ext}} = 3\left(\overline{\Delta G_b} - (m/6)\cdot\Delta\Delta G_{mis}\right)$.

As long as $\Delta G_b > (m/6)\cdot \Delta\Delta G_{mis} \approx 1\ k_BT$, the optimal extension is non-zero, i.e., RecA-mediated extension enhances sequence detection. *Inset*: The experimental (black lines) and the theoretical (blue band) limitations on $\Delta G_{ext}$ and $\Delta G_b$. The optimal values (blue band) (4) are calculated, for all possible competitors ($m$ = 1, 2, 3) and the range of measured values of $\Delta\Delta G_{mis}$. The resulting bands are parallel and overlapping, a direct outcome of the optimality of $\Delta G_t$.

## Polymerization

## Homologous recognition and strand exchange

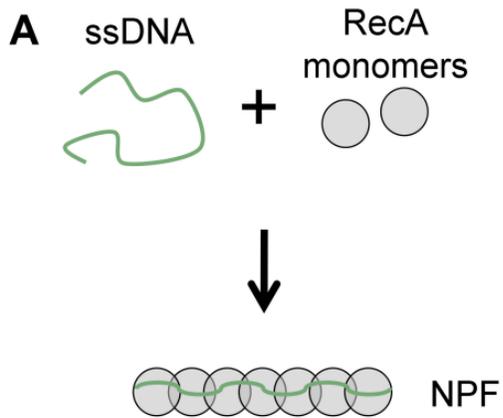
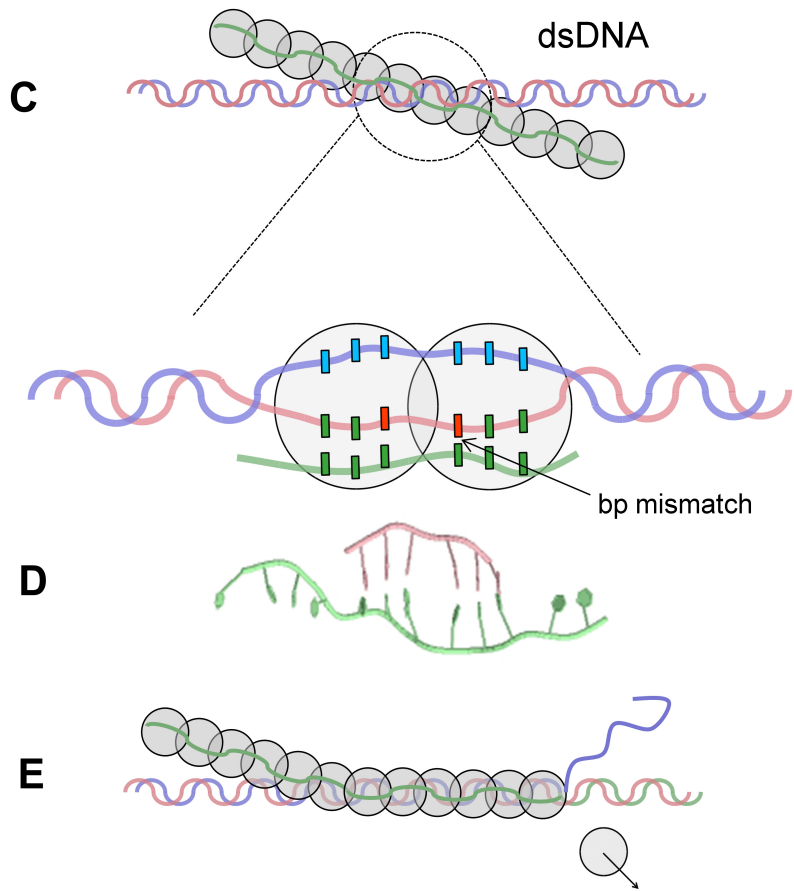

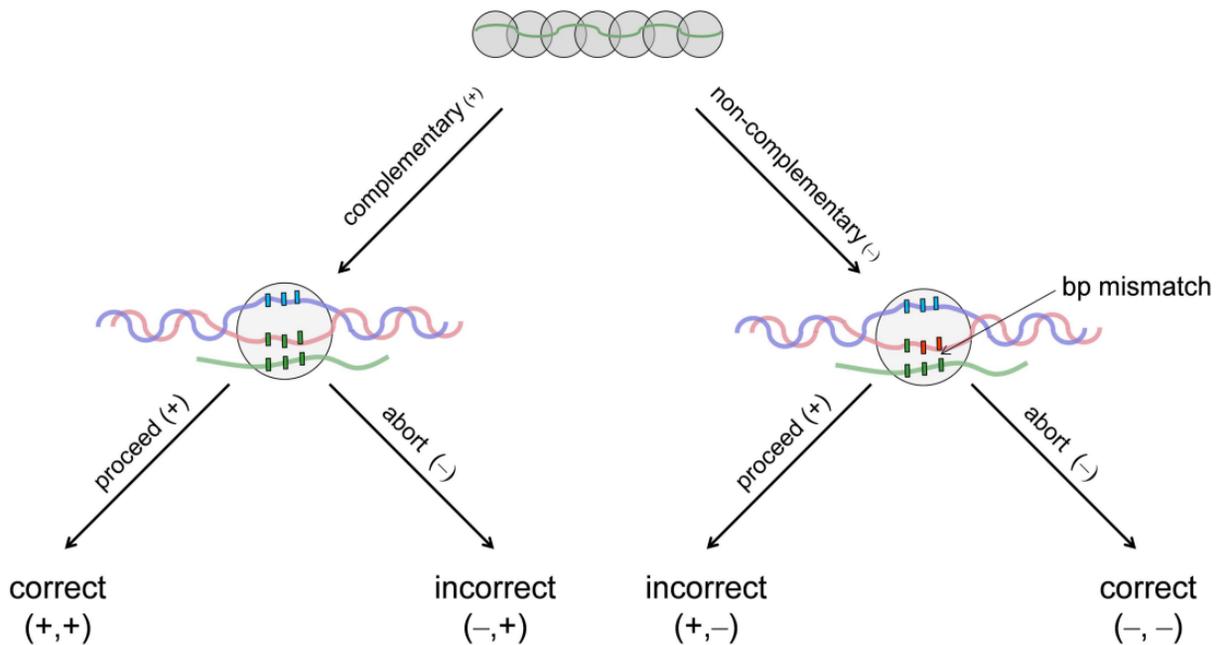

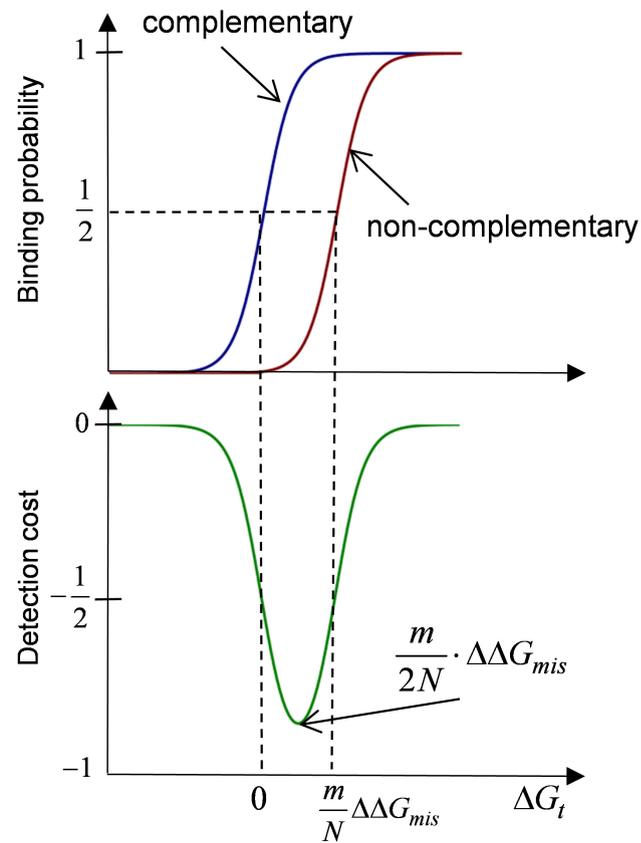
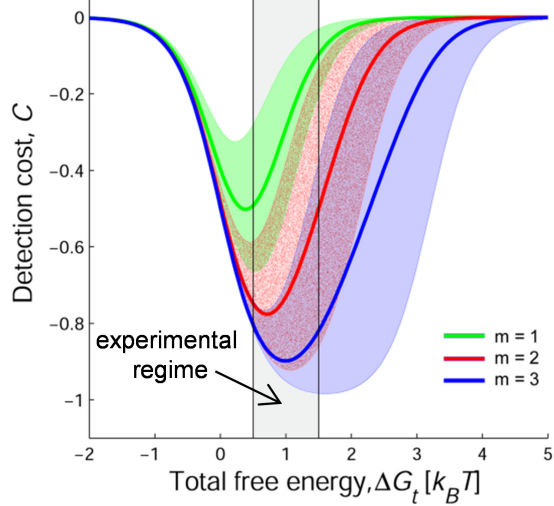
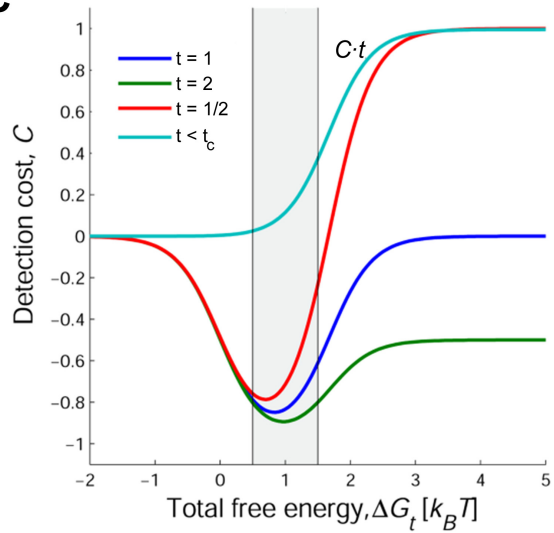

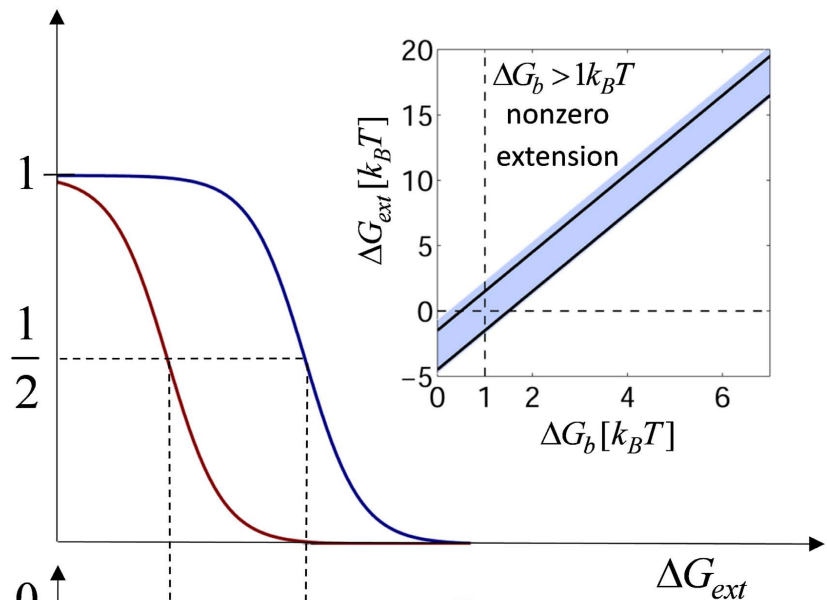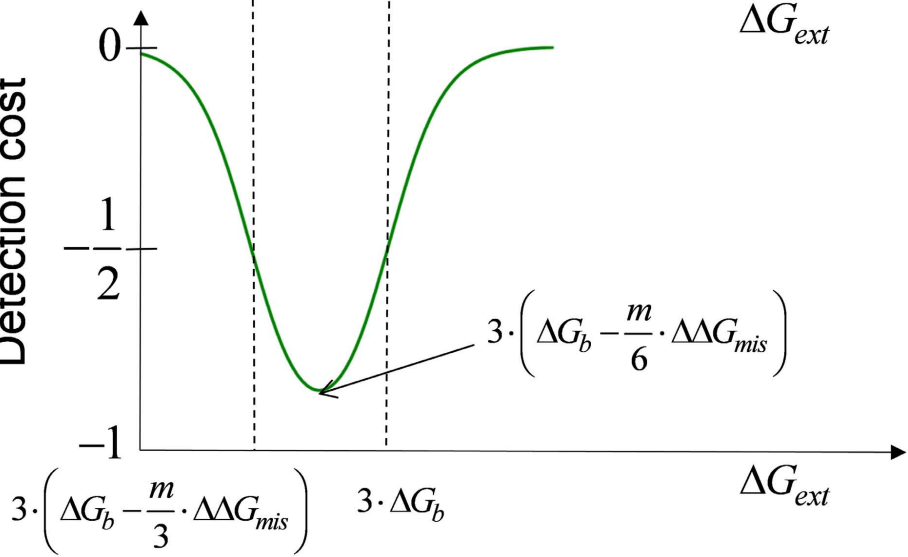